# On Multiple Scattering of Radiation by an Infinite Grating of Dielectric Circular Cylinders at Oblique Incidence

ÖMER KAVAKLIOĞLU and BARUCH SCHNEIDER

## Abstract

A rigorous analytical representation for the multiple scattering coefficients of the fields radiated by an infinite grating of dielectric circular cylinders excited by an obliquely incident plane electromagnetic wave is derived in terms of the "well-known scattering coefficients of an isolated dielectric cylinder at oblique incidence" and "Schlömilch series". In addition, a generalized sum-integral grating equation is acquired for the multiple scattered amplitude of a cylinder at oblique incidence in the grating in terms of the scattering coefficients of the insulating dielectric circular cylinder at oblique incidence.

---

## 1. Introduction

Twersky [1] treated the problem of multiple scattering of waves by an arbitrary configuration of parallel cylinders, and presented the formal solution in terms of cylindrical wave functions, taking into account all possible contributions to the excitation of a particular cylinder by the radiation scattered by the remaining cylinders. Expressing the scattered wave as an infinite sum of orders of scattering, he extended his solution to consider the case where all the axes of cylinders lie in the same plane [2], and applied the multiple scattering theories to the finite grating of cylinders [3]. He [4] exploited Green's function methods to represent the field of a grating of cylinders excited by a plane wave as sets of plane waves, and expressed the "multiple scattered amplitude of a cylinder in the grating" in terms of a scalar functional equation for the grating and the "single scattering amplitude of an isolated cylinder at normal incidence". He [5] then employed the separation-of-variables technique to obtain a set of algebraic equations for the "multiple scattering coefficients" in terms of the elementary function representations of "Schlömilch series" [6], and in terms of the "known scattering coefficients of an isolated cylinder at normal incidence". His results since then have been reiterated and extensively used by many other authors. For instance, Bogdanov *et al.* [7-10] have investigated various configurations of this classical diffraction problem. In a closely related area of acoustics, Cai and Williams [11-12] analysed the multiple scattering of anti-plane shear waves in fibre-reinforced composite materials by extending an exact analytical solution for a general two-dimensional multiple scattering problem to a formulation called as the "scatterer polymerization method" in order to reconstruct the solutions for abstract scatterers comprising lots of actual scatterers. Moreover, Cai [13] examined the "layered multiple scattering techniques" for anti-plane shear wave scattering from multiple gratings consisting of parallel cylinders.

The most generalized case of the diffraction by an infinite periodic array of perfectly conducting cylinders for obliquely incident plane electromagnetic waves, comprising the multiple scattering was first treated by Sivov [14], who determined the coefficients of reflection and transmission of an infinite grating in free space. In a more recent investigation for the "oblique case", Lee [15] inquired the "scattering of an obliquely incident electromagnetic wave by an arbitrary configuration of parallel, non-overlapping infinite cylinders" by constructing the expressions for the coefficients of the scattered wave potentials in terms of the independent scattering coefficients for an isolated cylinder, and presented the solution for the "scattering of an obliquely incident plane wave by a collection of closely-spaced, radially-stratified parallel cylinders that can have an arbitrary number of stratified layers" [16].

The purpose of this investigation is to acquire the *"exact analytical representation for the multiple scattering coefficients of an infinite grating of dielectric circular cylinders for obliquely incident plane electromagnetic waves"* in terms of the *"well-known scattering coefficients of an isolated dielectric circular cylinder at oblique incidence"*, which was originally derived by Wait [17] as long ago as 1955.

## 2. Problem formulation

*2.1. Multiple scattering representations of the fields: TM-mode*

We consider the scattering of plane electromagnetic waves by an infinite array of identical dielectric circular cylinders having infinite length with radius "$a$", dielectric constant "$\varepsilon_r$", and relative permeability "$\mu_r$". The cylinders of the grating are placed perpendicularly to the *x-y* plane along the $y-$axis, they are all parallel to the $z-$axis and separated by a distance of "$d$". A vertically polarized plane electromagnetic wave, which is obliquely incident upon the infinite grating of insulating dielectric circular cylinders, can be expanded in the cylindrical coordinate system $(R_s, \phi_s, z)$ of the $s^{th}$ cylinder in terms of the cylindrical waves referred to the axis of $s^{th}$ cylinder [17] as

$$\mathbf{E}_v^{inc}(R_s, \phi_s, z) = \hat{\mathbf{v}}_i E_{0v} e^{ik_r sd \sin\psi_i} \left\{ \sum_{n=-\infty}^{\infty} e^{-in\psi_i} J_n(k_r R_s) e^{in(\phi_s + \pi/2)} \right\} e^{-ik_z z} \tag{1}$$

In the above, $\hat{\mathbf{v}}_i$ denotes the vertical polarization vector associated with the unit vector having a component parallel to all the cylinders, $\phi_i$ is the angle of incidence in *x-y* plane measured from $x-$axis in such a way that $\psi_i = \pi + \phi_i$, implying that the wave is arbitrarily incident in the first quadrant of the coordinate system and "$J_n(x)$" stands for Bessel function of order *n*. In addition, we have

$$k_r = k_0 \sin\theta_i \tag{2a}$$

$$k_z = k_0 \cos\theta_i \tag{2b}$$

$\theta_i$ being the obliquity angle made with $z-$axis, and "$e^{-i\omega t}$" time dependence is suppressed throughout the paper, where "$\omega$" stands for the angular frequency of the incident wave in radians per second and "$t$" represents time in seconds.

*2.2. z-components of the exterior fields*

Denoting the location of the centres of the cylinders in the grating by $\mathbf{r}_0$, $\mathbf{r}_1$, $\mathbf{r}_2$, etc., the exact solution for the *z*-components of the electric field in the exterior of the grating can be expressed in terms of the incident electric field in the coordinate system of the $s^{th}$ cylinder located at $\mathbf{r}_s$, plus a summation of cylindrical waves outgoing from each of the individual $j^{th}$ cylinder located at $\mathbf{r}_j$, as $\left|\mathbf{r} - \mathbf{r}_j\right| \to \infty$, i. e.,

$$E_z^{(ext,TM)}(R_s,\phi_s,z) = E_z^{(inc,TM)}(R_s,\phi_s,z) + \sum_{j=-\infty}^{+\infty} E_z^{(j,TM)}(R_j,\phi_j,z) \tag{3}$$

Introducing an arbitrary oblique angle of arrival $\theta_i$ with the positive $z-$axis, the generalized exact representation of the multiple scattered fields by an infinite grating of dielectric cylinders for obliquely incident plane waves have been rigorously treated by the application of the method of separation-of-variables [18-20]. It has been proved that the $z-$component of the total electric and magnetic fields in the exterior of the infinite grating for *"obliquely incident and vertically polarized plane electromagnetic waves"* can be written [18] as

$$E_z^{(ext,TM)}(R_s,\phi_s,z) = \left\{ e^{ik_r sd\sin\psi_i} \sum_{n=-\infty}^{+\infty} \left[ \left( E_n^i + \sum_{m=-\infty}^{\infty} A_m^{(E,TM)} I_{n-m}(k_r d) \right) J_n(k_r R_s) + A_n^{(E,TM)} H_n^{(1)}(k_r R_s) \right] e^{in(\phi_s + \pi/2)} \right\} e^{-ik_z z} \tag{4a}$$

$$H_z^{(ext,TM)}(R_s,\phi_s,z) = \left\{ e^{ik_r sd\sin\psi_i} \sum_{n=-\infty}^{+\infty} \left[ \left( \sum_{m=-\infty}^{\infty} A_m^{(H,TM)} I_{n-m}(k_r d) \right) J_n(k_r R_s) + A_n^{(H,TM)} H_n^{(1)}(k_r R_s) \right] e^{in(\phi_s + \pi/2)} \right\} e^{-ik_z z} \tag{4b}$$

In the representation of the electric and magnetic fields above, $\left\{ A_n^{(E,TM)}, A_n^{(H,TM)} \right\}_{n=-\infty}^{\infty}$ denotes the set of all multiple scattering coefficients of the infinite grating associated with *"vertically polarized obliquely incident plane electromagnetic waves"*, $\forall n \ni n \in Z$, where *"Z"* represents the set of all integers. In expressions (4a, b), we have

$$E_n^i = \sin\theta_i E_{0v} e^{-in\psi_i} \tag{5a}$$

$$I_n(k_r d) = \sum_{p=1}^{+\infty} H_n^{(1)}(pk_r d) \left[ e^{ipk_r d\sin\psi_i}(-1)^n + e^{-ipk_r d\sin\psi_i} \right] \tag{5b}$$

and " $H_n^{(1)}(x)$ " denotes the $n^{th}$ order Hankel function of first kind. The series in expression (5b) is the well-known *"Schlömilch Series $I_{n-m}(k_r d)$ "* [6] and convergent when $k_r d(1 \pm \sin\psi_i)/2\pi$ does not equal integers.

*2.3. Exact Form of the Equations for the Multiple Scattering Coefficients $A_n^{(E,TM)}$ and $A_n^{(H,TM)}$ of the Infinite Grating at Oblique Incidence*

The generalized multiple scattering coefficients, $\left\{A_n^{(E,TM)}, A_n^{(H,TM)}\right\}_{n=-\infty}^{+\infty}$ of the infinite grating of dielectric circular cylinders excited by an obliquely incident vertically polarized plane wave, satisfy two infinite sets of equations in coupled form. In these infinite sets of equations, the undetermined coefficients of the electric fields $\left\{A_n^{(E,TM)}\right\}_{n=-\infty}^{+\infty}$ and magnetic fields $\left\{A_n^{(H,TM)}\right\}_{n=-\infty}^{+\infty}$ arise in coupled form. The rigorous derivation of these equations was published in [18].

These infinite sets of equations for the multiple scattering coefficients of the infinite grating, which has not been analytically solved yet, can be expressed in the form of a matrix equation in coupled form as

$$\begin{pmatrix} A_n^{(E,TM)} \\ A_n^{(H,TM)} \end{pmatrix} = \begin{pmatrix} b_n^{\mu} & 1 \\ 1 & -b_n^{\varepsilon} \end{pmatrix}^{-1} \begin{pmatrix} -b_n^{\mu} c_n & -a_n^{\mu} \\ -a_n^{\varepsilon} & b_n^{\varepsilon} c_n \end{pmatrix} \begin{pmatrix} E_n^i + \sum_{m=-\infty}^{\infty} A_m^{(E,TM)} I_{n-m}(k_r d) \\ \sum_{m=-\infty}^{\infty} A_m^{(H,TM)} I_{n-m}(k_r d) \end{pmatrix} \tag{6}$$

The element "$c_n$" in the matrix equation (6) can be written as

$$c_n := \frac{J_n(k_r a)}{H_n^{(1)}(k_r a)} \tag{7}$$

$\forall n \ni n \in Z$, and two sets of constants "$a_n^{\zeta}$" and "$b_n^{\zeta}$", in which $\zeta_r \in \{\varepsilon_r, \mu_r\}$ stands for the dielectric constant and the relative permeability of the dielectric cylinders respectively, are given as

$$a_n^{\zeta} := \frac{W_n^{JJ}(\zeta_r)}{W_n^{JH}(\zeta_r)} \tag{8a}$$

$$b_n^{\zeta} := \sqrt{\frac{\varepsilon_0 \mu_0}{\zeta_0^2}} \left[ \frac{J_n(k_1 a) H_n^{(1)}(k_r a)}{W_n^{JH}(\zeta_r)} \right] \left( \frac{inF}{k_r a} \right) \tag{8b}$$

$\forall n \ni n \in Z$, where "$k_1$" in (8a, b) is defined as $k_1 := k_0 \sqrt{\varepsilon_r \mu_r - \cos^2 \theta_i}$, with $\zeta \in \{\varepsilon, \mu\}$. In (8a, b), we have

$$W_n^{JJ}(\zeta_r) := J_n(k_1 a) J_n'(k_r a) - \zeta_r \left( \frac{k_r}{k_1} \right) J_n'(k_1 a) J_n(k_r a) \tag{9a}$$

$$W_n^{JH}(\zeta_r) := J_n(k_1 a) H_n^{(1)'}(k_r a) - \zeta_r \left( \frac{k_r}{k_1} \right) J_n'(k_1 a) H_n^{(1)}(k_r a) \tag{9b}$$

and *"F"* in expression (8b) is a constant which is given as

$$F := \frac{(\mu_r \varepsilon_r - 1)\cos\theta_i}{\mu_r \varepsilon_r - \cos^2\theta_i} \tag{10}$$

$J_n'(\varsigma)$ and $H_n^{(1)'}(\varsigma)$ in equation (9a, b) are defined as the first derivatives of these functions with respect to their arguments, i.e., $J_n'(\varsigma) \equiv \frac{d}{d\varsigma} J_n(\varsigma)$ and $H_n^{(1)'}(\varsigma) \equiv \frac{d}{d\varsigma} H_n^{(1)}(\varsigma)$. Setting $\zeta = \varepsilon, \mu$ in equation (8b), we have evaluated *"$b_n^\zeta$"* as

$$b_n^\varepsilon = i\xi_0 \left( \frac{nF}{k_r a} \right) \frac{J_n(k_1 a) H_n^{(1)}(k_r a)}{W_n^{JH}(\varepsilon_r)} \tag{11a}$$

$$b_n^\mu = i\eta_0 \left( \frac{nF}{k_r a} \right) \frac{J_n(k_1 a) H_n^{(1)}(k_r a)}{W_n^{JH}(\mu_r)} \tag{11b}$$

where $\xi_0$ and $\eta_0$ in (11a, b) are defined as

$$\xi_0 := \eta_0^{-1} := \sqrt{\frac{\mu_0}{\varepsilon_0}} \tag{12}$$

The product of the expressions in (11a, b) is a special one, and is given as

$$b_n^\varepsilon b_n^\mu = -\left( \frac{nF}{k_r a} \right)^2 \frac{[J_n(k_1 a) H_n^{(1)}(k_r a)]^2}{W_n^{JH}(\varepsilon_r) W_n^{JH}(\mu_r)} \tag{13}$$

$\forall n \ni n \in Z$. Inverting the matrix multiplying the unknown vector containing the multiple scattering coefficients associated with electric and magnetic fields in equation (6), we obtain a matrix equation for the unknown multiple scattering coefficients, $\left\{ A_n^{(E,TM)}, A_n^{(H,TM)} \right\}_{n=-\infty}^{+\infty}$, of the exterior electric and magnetic fields of the infinite grating as

$$\begin{pmatrix} A_n^{(E,TM)} \\ A_n^{(H,TM)} \end{pmatrix} = \begin{pmatrix} \Gamma_n^{\varepsilon\mu} & -\gamma_n^{\mu\varepsilon} \\ +\gamma_n^{\varepsilon\mu} & \Gamma_n^{\mu\varepsilon} \end{pmatrix} \begin{pmatrix} E_n^i + \sum_{m=-\infty}^{\infty} A_m^{(E,TM)} I_{n-m}(k_r d) \\ \sum_{m=-\infty}^{\infty} A_m^{(H,TM)} I_{n-m}(k_r d) \end{pmatrix} \tag{14}$$

After some algebraic manipulations the coefficients of the matrix in equation (14) can be obtained as

$$\Gamma_n^{\varepsilon\mu} = -\left( \frac{a_n^{\varepsilon} + b_n^{\varepsilon} b_n^{\mu} c_n}{1 + b_n^{\varepsilon} b_n^{\mu}} \right) \tag{15a}$$

$$\Gamma_n^{\mu\varepsilon} = -\left( \frac{a_n^{\mu} + b_n^{\varepsilon} b_n^{\mu} c_n}{1 + b_n^{\varepsilon} b_n^{\mu}} \right) \tag{15b}$$

for the diagonal elements, and

$$\gamma_n^{\varepsilon\mu} = +\left( \frac{b_n^{\mu} (a_n^{\varepsilon} - c_n)}{1 + b_n^{\varepsilon} b_n^{\mu}} \right) \tag{15c}$$

$$\gamma_n^{\mu\varepsilon} = +\left( \frac{b_n^{\varepsilon} (a_n^{\mu} - c_n)}{1 + b_n^{\varepsilon} b_n^{\mu}} \right) \tag{15d}$$

for the anti-diagonal elements.

**3. The Exact Analytical Solution for the Scattering Coefficients of the Infinite Grating at Oblique Incidence: Generalized Twersky's Representation**

*3.1. The Exact Matrix Form of the Equations for the Scattering Coefficients of the Infinite Grating at Oblique Incidence: Vertical Polarization*

Separating the matrix in equation (14) into two parts, we can express the undetermined multiple scattering coefficients of the infinite grating at oblique incidence as

$$\underbrace{\begin{pmatrix} A_n^{(E,TM)} \\ A_n^{(H,TM)} \end{pmatrix}}_{\substack{\text{The Scattering} \\ \text{Coefficients} \\ \text{of the Infinite} \\ \text{Grating at} \\ \text{Oblique Incidence}}} = \underbrace{\tilde{\mathbf{S}}_n \begin{pmatrix} E_n^i \\ 0 \end{pmatrix}}_{\substack{\text{The Single} \\ \text{Scattering} \\ \text{Approximation} \\ \text{at Oblique} \\ \text{Incidence}}} + \underbrace{\tilde{\mathbf{S}}_n \sum_{m=-\infty}^{\infty} \begin{pmatrix} A_m^{(E,TM)} \\ A_m^{(H,TM)} \end{pmatrix} I_{n-m}(k_r d)}_{\substack{\text{Contributions due to Multiple Scattering} \\ \text{of Radiation at Oblique Incidence}}} \tag{16}$$

In the coefficient equation above, $\tilde{\mathbf{S}}_n$ is defined as the *"normalized single-scattering coefficient matrix associated with an isolated dielectric circular cylinder at oblique incidence"* as

$$\tilde{\mathbf{S}}_n := \begin{pmatrix} \Gamma_n^{\varepsilon\mu} & -\gamma_n^{\mu\varepsilon} \\ +\gamma_n^{\varepsilon\mu} & \Gamma_n^{\mu\varepsilon} \end{pmatrix} \tag{17}$$

The equation (16) describes the multiple scattering coefficients for the exterior electric and magnetic fields of the infinite grating written in compact matrix form for obliquely incident vertically polarized waves. The first term on the right hand side of (16) which is a $2 \times 1$ vector, describes the solution for the scattering coefficients of the single dielectric cylinder at oblique incidence. This term is responsible for the single scattering by the infinite grating at oblique incidence. The second term, which is the *"normalized single-scattering coefficient matrix"* $\widetilde{\mathbf{S}}_n$, multiplied by a vector comprising the infinite sums of the scattering coefficients weighted by "*Schlömilch series*" accounts for the multiple scattering effects in the infinite grating. Since the multiple scattering effects will die out as the distance between the cylinders goes to infinity, the second term in equation (16) vanishes, yielding the solution for the scattering coefficients of an isolated infinitely long insulating dielectric circular cylinder at oblique incidence originally derived by Wait [17].

*3.2. Evaluation of the normalized single-scattering coefficient matrix* $\widetilde{\mathbf{S}}_n$

In this section, we will evaluate the elements of $\widetilde{\mathbf{S}}_n$, which emerge in the representation of the multiple scattering coefficients of the infinite grating for vertically polarized obliquely incident waves, defined explicitly in equation (14). Employing the expressions (7-13) in the evaluation of the first diagonal element of $\widetilde{\mathbf{S}}_n$, namely "$\Gamma_n^{\varepsilon\mu}$", in expression (15a), we have obtained

$$\Gamma_n^{\varepsilon\mu} = -\left\{ \frac{ \dfrac{W_n^{JJ}(\varepsilon_r)}{W_n^{JH}(\varepsilon_r)} - \left(\dfrac{nF}{k_r a}\right)^2 \left[ \dfrac{J_n^2(k_1 a) H_n^{(1)}(k_r a)}{W_n^{JH}(\varepsilon_r) W_n^{JH}(\mu_r)} \right] J_n(k_r a) }{ 1 - \left(\dfrac{nF}{k_r a}\right)^2 \dfrac{[J_n(k_1 a) H_n^{(1)}(k_r a)]^2}{W_n^{JH}(\varepsilon_r) W_n^{JH}(\mu_r)} } \right\} \tag{18}$$

$\forall n \ni n \in Z$. Inserting the expressions (9a, b) into the expression in (18), we can establish the relationship between "$\Gamma_n^{\varepsilon\mu}$" and "$a_{cyl,n}^{(E,TM)}$", the scattering coefficients of the exterior electric field of an infinitely long isolated circular dielectric cylinder associated with a vertically polarized obliquely incident wave as

$$\Gamma_n^{\varepsilon\mu} = \left\{ -\frac{J_n(k_r a)}{H_n^{(1)}(k_r a)} + \left(\frac{2i}{T\pi}\right) \frac{ \left[ \dfrac{H_n^{(1)'}(k_r a)}{k_r a H_n^{(1)}(k_r a)} - \dfrac{\mu_r J_n'(k_1 a)}{k_1 a J_n(k_1 a)} \right] }{ [k_r a H_n^{(1)}(k_r a)]^2 } \right\} = \frac{a_{cyl,n}^{(E,TM)}}{E_{0v} \sin\theta_i} := \widetilde{a}_{cyl,n}^{(E,TM)} \tag{19}$$

$\forall n \ni n \in Z$. The parameter "T" in expression (19) is given as

$$T := \left\{ \left[ \frac{H_n^{(1)'}(k_r a)}{k_r a H_n^{(1)}(k_r a)} - \frac{\varepsilon_r J_n'(k_1 a)}{k_1 a J_n(k_1 a)} \right] \left[ \frac{H_n^{(1)'}(k_r a)}{k_r a H_n^{(1)}(k_r a)} - \frac{\mu_r J_n'(k_1 a)}{k_1 a J_n(k_1 a)} \right] - n^2 \cos^2\theta_i \left( \frac{1}{(k_1 a)^2} - \frac{1}{(k_r a)^2} \right)^2 \right\} \tag{20}$$

The expression "$a_{cyl,n}^{(E,TM)}$" in (19) is nothing but Wait's result [17] for the scattering coefficients of an infinitely long isolated circular dielectric cylinder at oblique incidence for the exterior electric field associated with the vertical polarization case. The evaluation of the second diagonal element in $\widetilde{\mathbf{S}}_n$, namely "$\Gamma_n^{\mu\varepsilon}$" in expression (15b), yields

$$\Gamma_n^{\mu\varepsilon} = -\left\{ \frac{ \left[ \frac{W_n^{JJ}(\mu_r)}{W_n^{JH}(\mu_r)} - \left(\frac{nF}{k_r a}\right)^2 \left[ \frac{J_n^2(k_1 a) H_n^{(1)}(k_r a)}{W_n^{JH}(\varepsilon_r) W_n^{JH}(\mu_r)} \right] \right] J_n(k_r a) }{ 1 - \left(\frac{nF}{k_r a}\right)^2 \frac{[J_n(k_1 a) H_n^{(1)}(k_r a)]^2}{W_n^{JH}(\varepsilon_r) W_n^{JH}(\mu_r)} } \right\} \tag{21}$$

$\forall n \ni n \in Z$. Inserting expressions (9a, b) into expression in (21), we can establish the relationship between "$\Gamma_n^{\mu\varepsilon}$" and "$a_{cyl,n}^{(H,TE)}$", the scattering coefficients of the exterior magnetic field of an infinitely long isolated circular dielectric cylinder associated with a horizontally polarized obliquely incident wave as

$$\Gamma_n^{\mu\varepsilon} = \left\{ -\frac{J_n(k_r a)}{H_n^{(1)}(k_r a)} + \left(\frac{2i}{T\pi}\right) \frac{ \left[ \frac{H_n^{(1)\prime}(k_r a)}{k_r a H_n^{(1)}(k_r a)} - \frac{\varepsilon_r J_n'(k_1 a)}{k_1 a J_n(k_1 a)} \right] }{ [k_r a H_n^{(1)}(k_r a)]^2 } \right\} = \frac{a_{cyl,n}^{(H,TE)}}{H_{0v} \sin\theta_i} := \widetilde{a}_{cyl,n}^{(H,TE)} \tag{22}$$

The expression "$a_{cyl,n}^{(H,TE)}$" in (22) is identical to Wait's result [17] for the scattering coefficients an infinitely long isolated circular dielectric cylinder at oblique incidence associated with its exterior magnetic field corresponding to the horizontal polarization case. In expression (22), we have

$$H_n^i = \sin\theta_i H_{0v} e^{-in\psi_i} \tag{23a}$$

$$H_{0v} = \eta_0 E_{0h} \tag{23b}$$

On the other hand, the evaluation of the first anti-diagonal element of $\widetilde{\mathbf{S}}_n$, namely "$+\gamma_n^{\varepsilon\mu}$" in expression (15), yields

$$+\gamma_n^{\varepsilon\mu} = \left\{ \frac{ i\eta_0 \left(\frac{nF}{k_r a}\right) \frac{J_n(k_1 a) H_n^{(1)}(k_r a)}{W_n^{JH}(\mu_r)} \left[ \frac{W_n^{JJ}(\varepsilon_r)}{W_n^{JH}(\varepsilon_r)} - \frac{J_n(k_r a)}{H_n^{(1)}(k_r a)} \right] }{ 1 - \left(\frac{nF}{k_r a}\right)^2 \frac{[J_n(k_1 a) H_n^{(1)}(k_r a)]^2}{W_n^{JH}(\varepsilon_r) W_n^{JH}(\mu_r)} } \right\} \tag{24}$$

$\forall n \ni n \in Z$ . Upon inserting the expressions (9a, b) into the expression in (24), we can establish the relationship between " $+\gamma_n^{\varepsilon\mu}$ " and " $a_{cyl,n}^{(H,TM)}$ ", the scattering coefficients of the exterior magnetic field of an infinitely long isolated circular dielectric cylinder associated with a vertically polarized obliquely incident wave as

$$+\gamma_n^{\varepsilon\mu} = -\left\{ \frac{2\eta_0}{T\pi} \left( \frac{1}{(k_1 a)^2} - \frac{1}{(k_r a)^2} \right) \frac{n\cos\theta_i}{[k_r a H_n^{(1)}(k_r a)]^2} \right\} = \frac{a_{cyl,n}^{(H,TM)}}{E_{0v}\sin\theta_i} := \widetilde{a}_{cyl,n}^{(H,TM)} \tag{25}$$

$\forall n \ni n \in Z$ . The coefficient " $a_{cyl,n}^{(H,TM)}$ " in expression (25) is again Wait's result [17] for the scattering coefficients of the exterior magnetic field of an infinitely long isolated dielectric circular cylinder at oblique incidence associated with the vertical polarization case. The evaluation of the second anti-diagonal element in expression (16) for $\widetilde{\mathbf{S}}_n$ , namely, " $-\gamma_n^{\mu\varepsilon}$ " in equation (15d) yields

$$-\gamma_n^{\mu\varepsilon} = -\left\{ \frac{ i\xi_0 \left( \frac{nF}{k_r a} \right) \frac{J_n(k_1 a) H_n^{(1)}(k_r a)}{W_n^{JH}(\varepsilon_r)} \left[ \frac{W_n^{JJ}(\mu_r)}{W_n^{JH}(\mu_r)} - \frac{J_n(k_r a)}{H_n^{(1)}(k_r a)} \right] }{ 1 - \left( \frac{nF}{k_r a} \right)^2 \frac{[J_n(k_1 a) H_n^{(1)}(k_r a)]^2}{W_n^{JH}(\varepsilon_r) W_n^{JH}(\mu_r)} } \right\} \tag{26}$$

$\forall n \ni n \in Z$ . Employing the expressions (9a, b) in the expression in (26), we can establish the relationship between " $-\gamma_n^{\mu\varepsilon}$ " and " $a_{cyl,n}^{(E,TE)}$ ", the scattering coefficients of the exterior electric field of an infinitely long isolated circular dielectric cylinder associated with a horizontally polarized obliquely incident wave as

$$-\gamma_n^{\mu\varepsilon} = \left\{ \frac{2\xi_0}{T\pi} \left( \frac{1}{(k_1 a)^2} - \frac{1}{(k_r a)^2} \right) \frac{n\cos\theta_i}{[k_r a H_n^{(1)}(k_r a)]^2} \right\} = \frac{a_{cyl,n}^{(E,TE)}}{H_{0v}\sin\theta_i} := \widetilde{a}_{cyl,n}^{(E,TE)} \tag{27}$$

$\forall n \ni n \in Z$ . The coefficient " $a_{cyl,n}^{(E,TE)}$ " in expression (27) is again identical to Wait's [17] result for the scattering coefficients of the exterior electric field of an infinitely long isolated dielectric circular cylinder at oblique incidence corresponding to the horizontal polarization case. In addition, all the components of the fields in expressions (4a, b) reduce to those obtained by Wait [17] as the distance between the individual cylinders of the infinite grating approaches to infinity. The sign differences between Wait's solution and ours in (18-27) are due to the fact that we have adopted a time dependence of " $e^{-i\omega t}$ ", which in turn resulted a sign change for " $i$ ", and the scattering coefficients in this study are expressed in terms of $H_n^{(1)}(x)$ 's, Hankel functions of first kind and of order *n*. Inserting expressions (19, 22, 25, 27) into (17), we have acquired the simplified expression for $\widetilde{\mathbf{S}}_n$, as

$$\widetilde{\mathbf{S}}_n = \begin{pmatrix} a_{cyl,n}^{(E,TM)} & a_{cyl,n}^{(E,TE)} \\ a_{cyl,n}^{(H,TM)} & a_{cyl,n}^{(H,TE)} \end{pmatrix} \begin{pmatrix} \frac{1}{E_{0v}\sin\theta_i} & 0 \\ 0 & \frac{1}{H_{0v}\sin\theta_i} \end{pmatrix} \equiv \begin{pmatrix} \widetilde{a}_{cyl,n}^{(E,TM)} & \widetilde{a}_{cyl,n}^{(E,TE)} \\ \widetilde{a}_{cyl,n}^{(H,TM)} & \widetilde{a}_{cyl,n}^{(H,TE)} \end{pmatrix} := \widetilde{\mathbf{a}}_{\mathbf{n}} \tag{28}$$

It can easily be seen in (28), the matrix $\widetilde{\mathbf{S}}_n$ contains the *"normalized scattering coefficients of an infinitely long dielectric circular cylinder at oblique incidence for the exterior electric and magnetic fields associated with both polarizations"*. The scattering coefficients of an infinite circular dielectric cylinder at oblique incidence can then easily be expressed in terms of the *"normalized single scattering coefficient matrix,* $\widetilde{\mathbf{a}}_{\mathbf{n}}$ *"* in (28) as

$$\mathbf{a}_{\mathbf{n}} := \begin{pmatrix} a_{cyl,n}^{(E,TM)} & a_{cyl,n}^{(E,TE)} \\ a_{cyl,n}^{(H,TM)} & a_{cyl,n}^{(H,TE)} \end{pmatrix} = \begin{pmatrix} \widetilde{a}_{cyl,n}^{(E,TM)} & \widetilde{a}_{cyl,n}^{(E,TE)} \\ \widetilde{a}_{cyl,n}^{(H,TM)} & \widetilde{a}_{cyl,n}^{(H,TE)} \end{pmatrix} \begin{pmatrix} E_{0v}\sin\theta_i & 0 \\ 0 & H_{0v}\sin\theta_i \end{pmatrix} \tag{29}$$

In the above, $\mathbf{a}_{\mathbf{n}}$ represents the $2\times 2$ matrix associated with the *"single scattering coefficients of an infinitely long circular dielectric cylinder at oblique incidence"* acquired by Wait [17].

*3.3. Direct Neumann iteration procedure for the multiple scattering series representation of the scattering coefficients of the infinite grating at oblique incidence*

A representation in the form of multiple scattering series expansion for the scattering coefficients of the infinite grating at oblique incidence and its physical interpretation will be presented in this section. A direct Neumann iteration of equation (16) starting with the usual single scattering approximation as

$$\begin{pmatrix} A_{1,m}^{(E,TM)} \\ A_{1,m}^{(H,TM)} \end{pmatrix} := \begin{pmatrix} \Gamma_m^{\varepsilon\mu} \\ +\gamma_m^{\varepsilon\mu} \end{pmatrix} E_m^i \equiv \begin{pmatrix} a_{cyl,m}^{(E,TM)} \\ a_{cyl,m}^{(H,TM)} \end{pmatrix} e^{-im\psi_i} \tag{30}$$

yields the second order for the scattering coefficients of the electric and magnetic fields as

$$\underbrace{\begin{pmatrix} A_{2,n}^{(E,TM)} \\ A_{2,n}^{(H,TM)} \end{pmatrix}}_{\substack{\text{The Second Order} \\ \text{Multiple Scattering} \\ \text{Coefficients of the} \\ \text{InfiniteGrating at} \\ \text{Oblique Incidence}}} = \underbrace{\begin{pmatrix} \Gamma_n^{\varepsilon\mu} \\ +\gamma_n^{\varepsilon\mu} \end{pmatrix} E_n^i}_{\substack{\text{The First Order of} \\ \text{Scattering due to the} \\ \text{Excitation of Each} \\ \text{Cylinder by only} \\ \text{the Plane Wave}}} + \underbrace{\begin{pmatrix} \Gamma_n^{\varepsilon\mu} & -\gamma_n^{\mu\varepsilon} \\ +\gamma_n^{\varepsilon\mu} & \Gamma_n^{\mu\varepsilon} \end{pmatrix} \sum_{m-\infty}^{\infty} \begin{pmatrix} \Gamma_m^{\varepsilon\mu} \\ +\gamma_m^{\varepsilon\mu} \end{pmatrix} E_m^i \, l_{n-m}(k_r d)}_{\substack{\text{The Second Order of Scattering due to} \\ \text{the Excitation of Each Cylinder by} \\ \text{the First Order of Scattering from} \\ \text{the Remaining Cylinders at Oblique Incidence}}} \tag{31}$$

$\forall n \ni n \in Z$ . The first order consists of waves scattered by single cylinder, and the second order consists of waves scattered by two cylinders. Assuming this second order solution of (31) as our new guess in the following form

$$\begin{pmatrix} A_m^{(E,TM)} \\ A_m^{(H,TM)} \end{pmatrix} = \begin{pmatrix} \Gamma_m^{\varepsilon\mu} \\ +\gamma_m^{\varepsilon\mu} \end{pmatrix} E_m^i + \begin{pmatrix} \Gamma_m^{\varepsilon\mu} & -\gamma_m^{\mu\varepsilon} \\ +\gamma_m^{\varepsilon\mu} & \Gamma_m^{\mu\varepsilon} \end{pmatrix} \sum_{p-\infty}^{\infty} \begin{pmatrix} \Gamma_p^{\varepsilon\mu} \\ +\gamma_p^{\varepsilon\mu} \end{pmatrix} E_p^i \, I_{m-p}(k_r d) \qquad (32)$$

and inserting (32) into (16), we obtained the third order of TM multiple scattering coefficients as

$$\underbrace{\begin{pmatrix} A_{3,n}^{(E,TM)}(\psi_i) \\ A_{3,n}^{(H,TM)}(\psi_i) \end{pmatrix}}_{\substack{\text{The Third Order Multiple Scattering Coefficients} \\ \text{of the InfiniteGrating at Oblique Incidence}}} = \underbrace{\begin{pmatrix} a_{cyl,n}^{(E,TM)} \\ a_{cyl,n}^{(H,TM)} \end{pmatrix} e^{-in\psi_i}}_{\substack{\text{The First Order of Scattering due to the Excitation} \\ \text{of Each Cylinder by only the Plane Wave}}}$$

$$+ \underbrace{\begin{pmatrix} a_{cyl,n}^{(E,TM)} & a_{cyl,n}^{(E,TE)} \\ a_{cyl,n}^{(H,TM)} & a_{cyl,n}^{(H,TE)} \end{pmatrix} \begin{pmatrix} \frac{1}{E_{0v}\sin\theta_i} & 0 \\ 0 & \frac{1}{H_{0v}\sin\theta_i} \end{pmatrix} \sum_{m-\infty}^{\infty} \begin{pmatrix} a_{cyl,m}^{(E,TM)} \\ a_{cyl,m}^{(H,TM)} \end{pmatrix} e^{-im\psi_i} I_{n-m}(k_r d)}_{\substack{\text{The Second Order of Scattering due to the Excitation of Each Cylinder by} \\ \text{the First Order of Scattering from the Remaining Cylinders at Oblique Incidence}}} \qquad (33)$$

$$+ \underbrace{\begin{pmatrix} \Gamma_n^{\varepsilon\mu} & -\gamma_n^{\mu\varepsilon} \\ +\gamma_n^{\varepsilon\mu} & \Gamma_n^{\mu\varepsilon} \end{pmatrix} \sum_{m-\infty}^{\infty} \begin{pmatrix} \Gamma_m^{\varepsilon\mu} & -\gamma_m^{\mu\varepsilon} \\ +\gamma_m^{\varepsilon\mu} & \Gamma_m^{\mu\varepsilon} \end{pmatrix} \sum_{p=-\infty}^{\infty} \begin{pmatrix} \Gamma_p^{\varepsilon\mu} \\ +\gamma_p^{\varepsilon\mu} \end{pmatrix} E_p^i \, I_{m-p}(k_r d) \, I_{n-m}(k_r d)}_{\substack{\text{The Third Order of Scattering in terms of Sums of Products of Three "Normalized Single-Scattering Coefficient Matrices"} \\ \text{whose Elements Consist of the Scattering Coefficients of an Isolated Cylinder at Oblique Incidence}}}$$

Repeating the same iterative procedure, i.e. inserting (33) into (16), the fourth order of scattering is obtained as

$$\underbrace{\begin{pmatrix} A_{4,n}^{(E,TM)}(\psi_i) \\ A_{4,n}^{(H,TM)}(\psi_i) \end{pmatrix}}_{\substack{\text{The Fourth Order of Scattering} \\ \text{Coefficients of the Infinite Grating}}} = \underbrace{\begin{pmatrix} a_{cyl,n}^{(E,TM)} \\ a_{cyl,n}^{(H,TM)} \end{pmatrix} e^{-in\psi_i}}_{\substack{\text{The Scattering Coefficients of an} \\ \text{isolated cylinder at Oblique Incidence}}}$$

$$+ \underbrace{\begin{pmatrix} \tilde{a}_{cyl,n}^{(E,TM)} & \tilde{a}_{cyl,n}^{(E,TE)} \\ \tilde{a}_{cyl,n}^{(H,TM)} & \tilde{a}_{cyl,n}^{(H,TE)} \end{pmatrix} \begin{pmatrix} \sum_{m=-\infty}^{+\infty} a_{cyl,m}^{(E,TM)} e^{-im\psi_i} I_{n-m}(k_r d) \\ \sum_{m=-\infty}^{+\infty} a_{cyl,m}^{(H,TM)} e^{-im\psi_i} I_{n-m}(k_r d) \end{pmatrix}}_{\substack{\text{The Second Order of Scattering due to the Excitation of Each Cylinder by} \\ \text{the First Order of Scattering from the Remaining Cylinders at Oblique Incidence}}} \qquad (34)$$

$$+ \underbrace{\begin{pmatrix} \tilde{a}_{cyl,n}^{(E,TM)} & \tilde{a}_{cyl,n}^{(E,TE)} \\ \tilde{a}_{cyl,n}^{(H,TM)} & \tilde{a}_{cyl,n}^{(H,TE)} \end{pmatrix} \begin{pmatrix} \sum_{m=-\infty}^{+\infty} \tilde{a}_{cyl,m}^{(E,TM)} I_{n-m} & \sum_{m=-\infty}^{+\infty} \tilde{a}_{cyl,m}^{(E,TE)} I_{n-m} \\ \sum_{m=-\infty}^{+\infty} \tilde{a}_{cyl,m}^{(H,TM)} I_{n-m} & \sum_{m=-\infty}^{+\infty} \tilde{a}_{cyl,m}^{(H,TE)} I_{n-m} \end{pmatrix} \begin{pmatrix} \sum_{p=-\infty}^{+\infty} a_{cyl,p}^{(E,TM)} e^{-ip\psi_i} I_{m-p} \\ \sum_{p=-\infty}^{+\infty} a_{cyl,p}^{(H,TM)} e^{-ip\psi_i} I_{m-p} \end{pmatrix}}_{\substack{\text{The Third Order of Scattering in terms of Sums of Products of Three "Single-Scattering Coefficient Matrices"} \\ \text{whose Elements Consist of the Scatterring Coefficients of an Isolated Cylinder at Oblique Incidence}}}$$

$$+ \underbrace{\begin{pmatrix} \Gamma_n^{\varepsilon\mu} & -\gamma_n^{\mu\varepsilon} \\ +\gamma_n^{\varepsilon\mu} & \Gamma_n^{\mu\varepsilon} \end{pmatrix} \sum_{m=-\infty}^{\infty} \begin{pmatrix} \Gamma_m^{\varepsilon\mu} & -\gamma_m^{\mu\varepsilon} \\ +\gamma_m^{\varepsilon\mu} & \Gamma_m^{\mu\varepsilon} \end{pmatrix} \sum_{p=-\infty}^{\infty} \begin{pmatrix} \Gamma_p^{\varepsilon\mu} & -\gamma_p^{\mu\varepsilon} \\ +\gamma_p^{\varepsilon\mu} & \Gamma_p^{\mu\varepsilon} \end{pmatrix} \sum_{t=-\infty}^{\infty} \begin{pmatrix} \Gamma_t^{\varepsilon\mu} \\ +\gamma_t^{\varepsilon\mu} \end{pmatrix} E_t^i I_{p-t} I_{m-p} I_{n-m}}_{\substack{\text{The Fourth Order of Scattering in terms of Sums of Products of Four "Single-Scattering Coefficient Matrices"} \\ \text{whose Elements Consist of the Scatterring Coefficients of an Isolated Cylinder at Oblique Incidence}}}$$

*3.4. The exact form of the equations for the multiple scattering coefficients $A_n^{(H,TE)}$ and $A_n^{(E,TE)}$ of the infinite grating at oblique incidence: Horizontal polarization*

The $z$-components of the total magnetic and electric fields in the exterior of the infinite grating for *"obliquely incident and horizontally polarized plane electromagnetic waves"* are expressed in [20] as

$$H_z^{(ext,TE)}(R_s,\phi_s,z)=\left\{e^{ik_r sd\sin\psi_i}\sum_{n=-\infty}^{+\infty}\left[\left(H_n^i+\sum_{m=-\infty}^{\infty}A_m^{(H,TE)}I_{n-m}(k_r d)\right)J_n(k_r R_s)\right.\right.$$

$$\left.\left.+A_n^{(H,TE)}H_n^{(1)}(k_r R_s)\right]e^{in(\phi_s+\pi/2)}\right\}e^{-ik_z z} \quad (35a)$$

$$E_z^{(ext,TE)}(R_s,\phi_s,z)=\left\{e^{ik_r sd\sin\psi_i}\sum_{n=-\infty}^{+\infty}\left[\left(\sum_{m=-\infty}^{\infty}A_m^{(E,TE)}I_{n-m}(k_r d)\right)J_n(k_r R_s)\right.\right.$$

$$\left.\left.+A_n^{(E,TE)}H_n^{(1)}(k_r R_s)\right]e^{in(\phi_s+\pi/2)}\right\}e^{-ik_z z} \quad (35b)$$

$\forall n \ni n \in Z$ . Two infinite sets of undetermined multiple scattering coefficients for the exterior magnetic and electric fields of the infinite grating associated with the *"horizontally polarized obliquely incident waves"*, namely $\left\{A_n^{(H,TE)},A_n^{(E,TE)}\right\}_{n=-\infty}^{+\infty}$ in expressions (35a, b) are given as

$$\begin{pmatrix}A_n^{(H,TE)}\\ A_n^{(E,TE)}\end{pmatrix}=\begin{pmatrix}\Gamma_n^{\mu\varepsilon} & +\gamma_n^{\varepsilon\mu}\\ -\gamma_n^{\mu\varepsilon} & \Gamma_n^{\varepsilon\mu}\end{pmatrix}\begin{pmatrix}H_n^i+\sum_{m=-\infty}^{\infty}A_m^{(H,TE)}I_{n-m}(k_r d)\\ \sum_{m=-\infty}^{\infty}A_m^{(E,TE)}I_{n-m}(k_r d)\end{pmatrix} \quad (36a)$$

$\forall n \ni n \in Z$ . Interchanging the order of these equations, we can write

$$\begin{pmatrix}A_n^{(E,TE)}\\ A_n^{(H,TE)}\end{pmatrix}=\begin{pmatrix}\Gamma_n^{\varepsilon\mu} & -\gamma_n^{\mu\varepsilon}\\ +\gamma_n^{\varepsilon\mu} & \Gamma_n^{\mu\varepsilon}\end{pmatrix}\begin{pmatrix}\sum_{m=-\infty}^{\infty}A_m^{(E,TE)}I_{n-m}(k_r d)\\ H_n^i+\sum_{m=-\infty}^{\infty}A_m^{(H,TE)}I_{n-m}(k_r d)\end{pmatrix} \quad (36b)$$

$\forall n \ni n \in Z$ . Since the equation (36b) is structurally identical to (16), the analytic solutions previously obtained for TM-mode can easily be modified to generate an infinite series expansion for the TE-mode. Separating the matrix in equation (36b) into two parts, we can express the multiple scattering coefficients of the exterior electric and magnetic fields of the infinite grating for the *"horizontally polarized obliquely incident plane waves"* as

$$\underbrace{\begin{pmatrix}A_n^{(E,TE)}\\ A_n^{(H,TE)}\end{pmatrix}}_{\substack{\text{The Scattering}\\ \text{Coefficients}\\ \text{of the Infinite}\\ \text{Grating at}\\ \text{Oblique Incidence}}}=\underbrace{\tilde{\mathbf{S}}_n\begin{pmatrix}0\\ H_n^i\end{pmatrix}}_{\substack{\text{The Single}\\ \text{Scattering}\\ \text{Approximation}\\ \text{at Oblique}\\ \text{Incidence}}}+\underbrace{\tilde{\mathbf{S}}_n\sum_{m=-\infty}^{\infty}\begin{pmatrix}A_m^{(E,TE)}\\ A_m^{(H,TE)}\end{pmatrix}I_{n-m}(k_r d)}_{\substack{\text{Contributions due to Multiple Scattering}\\ \text{of Radiation at Oblique Incidence}}} \quad (37)$$

*3.5. Neumann Series representation of the multiple scattering coefficients of the infinite grating at oblique incidence: Horizontal polarization*

The equations for the TE multiple scattering coefficients of the infinite grating at oblique incidence in (37) is the "dual" of the equations (16) for the TM mode. Therefore, the approximate solution for the scattering coefficients which is valid up to the fourth order could be obtained by a similar procedure as

$$\underbrace{\begin{pmatrix} A_{4,n}^{(E,TE)}(\psi_i) \\ A_{4,n}^{(H,TE)}(\psi_i) \end{pmatrix}}_{\substack{\text{The Fourth Order of Scattering} \\ \text{Coefficients of the Infinite Grating}}} = \underbrace{\begin{pmatrix} a_{cyl,n}^{(E,TE)} \\ a_{cyl,n}^{(H,TE)} \end{pmatrix} e^{-in\psi_i}}_{\substack{\text{The Scattering Coefficients of an} \\ \text{isolated cylinder at Oblique Incidence}}}$$

$$+\underbrace{\begin{pmatrix} \tilde{a}_{cyl,n}^{(E,TM)} & \tilde{a}_{cyl,n}^{(E,TE)} \\ \tilde{a}_{cyl,n}^{(H,TM)} & \tilde{a}_{cyl,n}^{(H,TE)} \end{pmatrix} \begin{pmatrix} \sum_{m=-\infty}^{+\infty} a_{cyl,m}^{(E,TE)} e^{-im\psi_i} \mathsf{l}_{n-m}(k_r d) \\ \sum_{m=-\infty}^{+\infty} a_{cyl,m}^{(H,TE)} e^{-im\psi_i} \mathsf{l}_{n-m}(k_r d) \end{pmatrix}}_{\substack{\text{The Second Order of Scattering due to the Excitation of Each Cylinder by} \\ \text{the First Order of Scattering from the Remaining Cylinders at Oblique Incidence}}} \tag{38}$$

$$+\underbrace{\begin{pmatrix} \tilde{a}_{cyl,n}^{(E,TM)} & \tilde{a}_{cyl,n}^{(E,TE)} \\ \tilde{a}_{cyl,n}^{(H,TM)} & \tilde{a}_{cyl,n}^{(H,TE)} \end{pmatrix} \begin{pmatrix} \sum_{m=-\infty}^{+\infty} \tilde{a}_{cyl,m}^{(E,TM)} \mathsf{l}_{n-m} & \sum_{m=-\infty}^{+\infty} \tilde{a}_{cyl,m}^{(E,TE)} \mathsf{l}_{n-m} \\ \sum_{m=-\infty}^{+\infty} \tilde{a}_{cyl,m}^{(H,TM)} \mathsf{l}_{n-m} & \sum_{m=-\infty}^{+\infty} \tilde{a}_{cyl,m}^{(H,TE)} \mathsf{l}_{n-m} \end{pmatrix} \begin{pmatrix} \sum_{p=-\infty}^{+\infty} a_{cyl,p}^{(E,TE)} e^{-ip\psi_i} \mathsf{l}_{m-p} \\ \sum_{p=-\infty}^{+\infty} a_{cyl,p}^{(H,TE)} e^{-ip\psi_i} \mathsf{l}_{m-p} \end{pmatrix}}_{\substack{\text{The Third Order of Scattering in terms of Sums of Products of Three "Single-Scattering Coefficient Matrices"} \\ \text{whose Elements Consist of the Scatterring Coefficients of an Isolated Cylinder at Oblique Incidence}}}$$

$$+\underbrace{\begin{pmatrix} \Gamma_n^{\varepsilon\mu} & -\gamma_n^{\mu\varepsilon} \\ +\gamma_n^{\varepsilon\mu} & \Gamma_n^{\mu\varepsilon} \end{pmatrix} \sum_{m=-\infty}^{\infty} \begin{pmatrix} \Gamma_m^{\varepsilon\mu} & -\gamma_m^{\mu\varepsilon} \\ +\gamma_m^{\varepsilon\mu} & \Gamma_m^{\mu\varepsilon} \end{pmatrix} \sum_{p=-\infty}^{\infty} \begin{pmatrix} \Gamma_p^{\varepsilon\mu} & -\gamma_p^{\mu\varepsilon} \\ +\gamma_p^{\varepsilon\mu} & \Gamma_p^{\mu\varepsilon} \end{pmatrix} \sum_{t=-\infty}^{\infty} \begin{pmatrix} -\gamma_t^{\mu\varepsilon} \\ \Gamma_t^{\mu\varepsilon} \end{pmatrix} H_t^i \mathsf{l}_{p-t} \mathsf{l}_{m-p} \mathsf{l}_{n-m}}_{\substack{\text{The Fourth Order of Scattering in terms of Sums of Products of Four "Single-Scattering Coefficient Matrices"} \\ \text{whose Elements Consist of the Scatterring Coefficients of an Isolated Cylinder at Oblique Incidence}}}$$

**4. The Functional Vector-Matrix Equation for the Multiple Scattering Amplitudes of the Infinite Grating at Oblique Incidence: Generalized Twersky's Representation**

*4.1. The matrix grating equation for the normalized multiple scattering coefficients of the infinite grating at oblique incidence*

The equations for the multiple scattering coefficients associated with the TM-mode can be rewritten in matrix form as

$$\mathbf{A}_n^{(\chi)}(\psi_i) = \tilde{\mathbf{S}}_n \left[ \mathbf{1}_g^{(\chi)} e^{-in\psi_i} + \sum_{m=-\infty}^{\infty} \mathbf{A}_m^{(\chi)}(\psi_i) \mathsf{l}_{n-m}(k_r d) \right] \tag{39a}$$

$$\mathbf{A}_n^{(\chi)}(\psi_i) = \begin{pmatrix} A_n^{(E,\chi)}(\psi_i) \\ A_n^{(H,\chi)}(\psi_i) \end{pmatrix}, \quad \chi = TM, TE \qquad (2\times 1) \text{ vector} \tag{39b}$$

$$\mathbf{1}_g^{(TM)} = \mathbf{1}_g^{(\uparrow)} E_{0v} \sin\theta_i \, ; \qquad \mathbf{1}_g^{(\uparrow)} := \begin{pmatrix} 1 \\ 0 \end{pmatrix} \qquad (2\times 1) \text{ vector} \tag{39c}$$

$$\mathbf{1}_{\mathrm{g}}^{(TE)} = \mathbf{1}_{\mathrm{g}}^{(\downarrow)} H_{0v} \sin\theta_i ; \qquad \mathbf{1}_{\mathrm{g}}^{(\downarrow)} := \begin{pmatrix} 0 \\ 1 \end{pmatrix} \quad (2\times 1) \text{ vector} \tag{39d}$$

$\forall n \ni n \in Z$ . In the equation above, $\mathbf{A}_n^{(\chi)}(\psi_i)$ , $\forall \chi \ni \chi = \{TM, TE\}$ , is a $2\times 1$ vector comprising the multiple scattering coefficients of the infinite grating for obliquely incident waves for the scattered electric and magnetic fields associated with both vertically polarized waves for $\chi \ni \chi = \{TM\}$ , that is the *"transverse-magnetic mode"*, and horizontally polarized waves for $\chi \ni \chi = \{TE\}$ , which denotes the *"transverse-electric mode"*. Using (39c) in (39a), we can rewrite equation (39a) as

$$\mathbf{A}_n^{(TM)}(\psi_i) = \widetilde{\mathbf{S}}_n \left[ \mathbf{1}_{\mathrm{g}}^{(\uparrow)} e^{-in\psi_i} E_{0v} \sin\theta_i + \sum_{m=-\infty}^{\infty} \mathbf{A}_m^{(TM)}(\psi_i) \mathbf{I}_{n-m}(k_r d) \right] \tag{40a}$$

$$\mathbf{A}_n^{(TE)}(\psi_i) = \widetilde{\mathbf{S}}_n \left[ \mathbf{1}_{\mathrm{g}}^{(\downarrow)} e^{-in\psi_i} H_{0v} \sin\theta_i + \sum_{m=-\infty}^{\infty} \mathbf{A}_m^{(TE)}(\psi_i) \mathbf{I}_{n-m}(k_r d) \right] \tag{40b}$$

Simplification of the equations (40a, b) can be achieved by introducing

$$\mathbf{A}_n^{(TM)}(\psi_i) = \widetilde{\mathbf{A}}_n^{(TM)}(\psi_i) E_{0v} \sin\theta_i \tag{41a}$$

$$\mathbf{A}_n^{(TE)}(\psi_i) = \widetilde{\mathbf{A}}_n^{(TE)}(\psi_i) H_{0v} \sin\theta_i \tag{41b}$$

where $\widetilde{\mathbf{A}}_n^{(\chi)}(\psi_i)$ , $\forall \chi \ni \chi = \{TM, TE\}$ , stands for the normalized multiple scattering coefficients of the infinite grating for the *z*-components of the exterior electric and magnetic fields, associated with both *"vertically polarized"* and *"horizontally polarized"* obliquely incident waves, respectively. Employing (41a, b) in equation (40a, b), we have acquired the equations for the *"normalized multiple scattering coefficients of the infinite grating associated with both vertically and horizontally polarized obliquely incident waves"* as

$$\widetilde{\mathbf{A}}_n^{(TM)}(\psi_i) = \widetilde{\mathbf{S}}_n \left[ \mathbf{1}_{\mathrm{g}}^{(\uparrow)} e^{-in\psi_i} + \sum_{m=-\infty}^{\infty} \widetilde{\mathbf{A}}_m^{(TM)}(\psi_i) \mathbf{I}_{n-m}(k_r d) \right] \tag{42a}$$

$$\widetilde{\mathbf{A}}_n^{(TE)}(\psi_i) = \widetilde{\mathbf{S}}_n \left[ \mathbf{1}_{\mathrm{g}}^{(\downarrow)} e^{-in\psi_i} + \sum_{m=-\infty}^{\infty} \widetilde{\mathbf{A}}_m^{(TE)}(\psi_i) \mathbf{I}_{n-m}(k_r d) \right] \tag{42b}$$

In order to be able to combine the equations (42a, b) into a unique one, we introduce the *"normalized multiple scattering coefficient matrix of the infinite grating for obliquely incident waves"*, $\widetilde{\mathbf{A}}_{\mathbf{n}}(\psi_{\mathbf{i}})$ , which is a $2\times 2$ matrix containing the multiple scattering coefficients of the infinite grating for the scattered electric and magnetic fields in such a way that those associated with *"vertically polarized* *obliquely incident waves"*, namely $\widetilde{\mathbf{A}}_n^{(TM)}(\psi_i)$, are placed in its first column, and those associated with *"horizontally polarized obliquely incident waves"*, namely $\widetilde{\mathbf{A}}_n^{(TE)}(\psi_i)$, are placed in its second column, as

$$\widetilde{\mathbf{A}}_{\mathbf{n}}(\boldsymbol{\psi}_{\mathbf{i}}) := \left(\widetilde{\mathbf{A}}_n^{(TM)}(\psi_i) \quad \widetilde{\mathbf{A}}_n^{(TE)}(\psi_i)\right) \equiv \begin{pmatrix} \widetilde{A}_n^{(E,TM)}(\psi_i) & \widetilde{A}_n^{(E,TE)}(\psi_i) \\ \widetilde{A}_n^{(H,TM)}(\psi_i) & \widetilde{A}_n^{(H,TE)}(\psi_i) \end{pmatrix} \tag{43}$$

In terms of this multiple scattering coefficient matrix, the equations for the multiple scattering coefficients defined in (40a, b) can be combined into a single matrix equation as

$$\left(\widetilde{\mathbf{A}}_n^{(TM)}(\psi_i) \quad \widetilde{\mathbf{A}}_n^{(TE)}(\psi_i)\right) = \widetilde{\mathbf{S}}_n \left[ \left(\mathbf{1}_{\mathbf{g}}^{(\uparrow)} \quad \mathbf{1}_{\mathbf{g}}^{(\downarrow)}\right) e^{-in\psi_i} + \sum_{m=-\infty}^{\infty} \left(\widetilde{\mathbf{A}}_m^{(TM)}(\psi_i) \quad \widetilde{\mathbf{A}}_m^{(TE)}(\psi_i)\right) \mathbf{I}_{n-m}(k_r d) \right] \tag{44}$$

Recognizing $\left(\mathbf{1}_{\mathbf{g}}^{(\uparrow)} \quad \mathbf{1}_{\mathbf{g}}^{(\downarrow)}\right) := \mathbf{1}$, that is the $2 \times 2$ identity matrix, we have acquired a *"generalized matrix equation for the normalized multiple scattering coefficients of the infinite grating"* associated with both vertically and horizontally polarized obliquely incident waves as

$$\widetilde{\mathbf{A}}_{\mathbf{n}}(\boldsymbol{\psi}_{\mathbf{i}}) = \widetilde{\mathbf{a}}_{\mathbf{n}} \left[ e^{-in\psi_i} + \sum_{m=-\infty}^{\infty} \widetilde{\mathbf{A}}_{\mathbf{m}}(\boldsymbol{\psi}_{\mathbf{i}}) \mathbf{I}_{n-m}(k_r d) \right] \tag{45}$$

This is the generalization of the *"Twersky's scalar grating equation"* to the case of obliquely incident waves, which was originally derived by Twersky [5] for the *"non-oblique incidence"* case.

*4.2. The functional vector equation for the multiple scattering amplitudes of the infinite grating at oblique incidence*

In order to express the multiple scattered amplitudes of the infinite grating at oblique incidence in terms of the single scattering amplitudes of an isolated cylinder, we have introduced the definition of the *"normalized single scattering amplitude matrix,"* $\widetilde{\mathbf{g}}_{\mathbf{z}}(\boldsymbol{\varphi}, \boldsymbol{\psi}_{\mathbf{i}})$, associated with the *"scattering amplitudes of an isolated infinitely long dielectric circular cylinder at oblique incidence"* as

$$\widetilde{\mathbf{g}}_{\mathbf{z}}(\boldsymbol{\varphi}, \boldsymbol{\psi}_{\mathbf{i}}) = \begin{pmatrix} g_\Gamma^{\varepsilon\mu}(\phi, \psi_i) & -g_\gamma^{\mu\varepsilon}(\phi, \psi_i) \\ +g_\gamma^{\varepsilon\mu}(\phi, \psi_i) & g_\Gamma^{\mu\varepsilon}(\phi, \psi_i) \end{pmatrix} := \sum_{n=-\infty}^{\infty} \begin{pmatrix} \Gamma_n^{\varepsilon\mu} & -\gamma_n^{\mu\varepsilon} \\ +\gamma_n^{\varepsilon\mu} & \Gamma_n^{\mu\varepsilon} \end{pmatrix} e^{in(\phi-\psi_i)} \tag{46}$$

and using (29) in (46), we can write

$$\widetilde{\mathbf{g}}_{\mathbf{z}}(\boldsymbol{\varphi}, \boldsymbol{\psi}_{\mathbf{i}}) := \sum_{n=-\infty}^{\infty} \widetilde{\mathbf{a}}_{\mathbf{n}} e^{in(\phi-\psi_i)} = \begin{pmatrix} \sum_{n=-\infty}^{\infty} \widetilde{a}_{cyl,n}^{(E,TM)} e^{in(\phi-\psi_i)} & \sum_{n=-\infty}^{\infty} \widetilde{a}_{cyl,n}^{(E,TE)} e^{in(\phi-\psi_i)} \\ \sum_{n=-\infty}^{\infty} \widetilde{a}_{cyl,n}^{(H,TM)} e^{in(\phi-\psi_i)} & \sum_{n=-\infty}^{\infty} \widetilde{a}_{cyl,n}^{(H,TE)} e^{in(\phi-\psi_i)} \end{pmatrix} \tag{47}$$

In the above, $\tilde{\mathbf{g}}_{\mathbf{z}}(\varphi,\psi_i)$ is a $2\times 2$ matrix and its elements contains the normalized scattering amplitudes of an isolated cylinder at oblique incidence for the exterior electric and magnetic fields associated with both vertically and horizontally polarized waves. Comparing (46) and (47), we conclude that $g_{\Gamma}^{\varepsilon\mu}(\phi,\psi_i)$ and $+g_{\gamma}^{\varepsilon\mu}(\phi,\psi_i)$ are the scattering amplitudes of an isolated dielectric cylinder for *"obliquely incident and vertically polarized"* plane waves associated with the *z*-components of the exterior electric and magnetic fields, respectively. On the other hand, $g_{\Gamma}^{\mu\varepsilon}(\phi,\psi_i)$ and $-g_{\gamma}^{\mu\varepsilon}(\phi,\psi_i)$ delineate the single scattering amplitudes of the same cylinder for *"obliquely incident and horizontally polarized"* plane waves associated with the *z*-components of the exterior magnetic and electric fields, respectively. In this depiction, $g_{\Gamma}^{\varepsilon\mu}(\phi,\psi_i)$ and $g_{\Gamma}^{\mu\varepsilon}(\phi,\psi_i)$ are the *"dual"* of each other, i.e., the scattering amplitude of an isolated dielectric circular cylinder for the *z*-component of its exterior magnetic field, namely $g_{\Gamma}^{\mu\varepsilon}(\phi,\psi_i)$, associated with the *"horizontally polarized"* and obliquely incident plane electromagnetic wave is the dual of the scattering amplitude of the same cylinder for the *z*-component of its exterior scattered electric field, $g_{\Gamma}^{\varepsilon\mu}(\phi,\psi_i)$, associated with the *"vertically polarized"* and obliquely incident waves. Besides, $+g_{\gamma}^{\varepsilon\mu}(\phi,\psi_i)$ and $-g_{\gamma}^{\mu\varepsilon}(\phi,\psi_i)$ are the *"dual"* of each other as well, i.e., the scattering amplitude of an isolated dielectric cylinder for the *z*-component of its exterior scattered electric field, viz. $-g_{\gamma}^{\mu\varepsilon}(\phi,\psi_i)$, associated with the *"horizontally polarized"* and obliquely incident waves, is the dual of the scattering amplitude of the same cylinder for the *z*-component of its scattered magnetic field, that is, $+g_{\gamma}^{\varepsilon\mu}(\phi,\psi_i)$, associated with *"vertically polarized"* and obliquely incident waves. In order to be able to establish the connection with the multiple scattering amplitudes of the infinite grating, we have now defined the normalized vector multiple scattering amplitudes of the infinite grating $\tilde{\mathbf{G}}_z^{(\chi)}(\phi,\psi_i)$, as

$$\tilde{\mathbf{G}}_z^{(\chi)}(\phi,\psi_i)=\sum_{n=-\infty}^{\infty}\tilde{\mathbf{A}}_n^{(\chi)}(\psi_i)e^{in\phi} \qquad \forall\chi \ni \chi=\{TM,TE\}, \tag{48a}$$

$$\tilde{\mathbf{G}}_z^{(\chi)}(\phi,\psi_i):=\begin{pmatrix}\tilde{G}_z^{(E,\chi)}(\phi,\psi_i)\\ \tilde{G}_z^{(H,\chi)}(\phi,\psi_i)\end{pmatrix}\equiv\begin{pmatrix}\sum_{n=-\infty}^{\infty}\tilde{A}_n^{(E,\chi)}(\psi_i)e^{in\phi}\\ \sum_{n=-\infty}^{\infty}\tilde{A}_n^{(H,\chi)}(\psi_i)e^{in\phi}\end{pmatrix} \tag{48b}$$

In the definition above, $\tilde{\mathbf{G}}_z^{(\chi)}(\phi,\psi_i)$ depicts the normalized vector multiple scattering amplitude of the infinite grating corresponding to the *"vertically polarized"* for $\chi \ni \chi=\{TM\}$, and to the *"horizontally* *polarized"* for $\chi \ni \chi = \{TE\}$, obliquely incident waves associated with the z-components of the electric and magnetic fields, respectively.

**Theorem 1:** *The relationship between the transverse-magnetic multiple scattering amplitudes of an infinite grating of dielectric circular cylinders (associated with its scattered electric and magnetic fields) for obliquely incident and vertically polarized plane waves satisfies the following vector-matrix functional equation in which the multiple scattering amplitudes of the infinite grating associated with the scattered electric and magnetic fields appear in coupled form. Furthermore, the single-scattering amplitudes associated with the scattered electric and magnetic fields of an isolated infinitely long cylinder corresponding to both "vertically and horizontally polarized obliquely incident waves" arise together in coupled form in this functional vector equation.*

$$\widetilde{\mathbf{G}}_z^{(TM)}(\phi,\psi_i) = \widetilde{\mathbf{g}}_{\mathbf{z}}(\boldsymbol{\varphi},\boldsymbol{\psi}_{\mathbf{i}})\mathbf{1}_{\mathbf{g}}^{(\uparrow)} + \boldsymbol{S}\, L_\varsigma \left[\widetilde{\mathbf{g}}_{\mathbf{z}}(\boldsymbol{\varphi},\boldsymbol{\varphi}_\varsigma)\widetilde{\mathbf{G}}_z^{(TM)}(\phi_\varsigma,\psi_i) + \widetilde{\mathbf{g}}_{\mathbf{z}}(\boldsymbol{\varphi},\boldsymbol{\pi}-\boldsymbol{\varphi}_\varsigma)\widetilde{\mathbf{G}}_z^{(TM)}(\pi-\phi_\varsigma,\psi_i)\right] \tag{49}$$

This the generalization of the Twersky's functional equation [4] for the multiple scattering amplitudes of the exterior electric and magnetic fields of the infinite grating of dielectric circular cylinders for *"vertically polarized"* obliquely incident waves, which can be written explicitly as

$$\begin{pmatrix} \widetilde{G}_z^{(E,TM)}(\phi,\psi_i) \\ \widetilde{G}_z^{(H,TM)}(\phi,\psi_i) \end{pmatrix} = \begin{pmatrix} g_\Gamma^{\varepsilon\mu}(\phi,\psi_i) \\ +g_\gamma^{\varepsilon\mu}(\phi,\psi_i) \end{pmatrix} + \boldsymbol{S}\, L_\varsigma \begin{pmatrix} g_\Gamma^{\varepsilon\mu}(\phi,\phi_\varsigma) & -g_\gamma^{\mu\varepsilon}(\phi,\phi_\varsigma) \\ +g_\gamma^{\varepsilon\mu}(\phi,\phi_\varsigma) & g_\Gamma^{\mu\varepsilon}(\phi,\phi_\varsigma) \end{pmatrix} \begin{pmatrix} \widetilde{G}_z^{(E,TM)}(\phi_\varsigma,\psi_i) \\ \widetilde{G}_z^{(H,TM)}(\phi_\varsigma,\psi_i) \end{pmatrix}$$
$$+ \boldsymbol{S}\, L_\varsigma \begin{pmatrix} g_\Gamma^{\varepsilon\mu}(\phi,\pi-\phi_\varsigma) & -g_\gamma^{\mu\varepsilon}(\phi,\pi-\phi_\varsigma) \\ +g_\gamma^{\varepsilon\mu}(\phi,\pi-\phi_\varsigma) & g_\Gamma^{\mu\varepsilon}(\phi,\pi-\phi_\varsigma) \end{pmatrix} \begin{pmatrix} \widetilde{G}_z^{(E,TM)}(\pi-\phi_\varsigma,\psi_i) \\ \widetilde{G}_z^{(H,TM)}(\pi-\phi_\varsigma,\psi_i) \end{pmatrix} \tag{50}$$

In the functional vector equation above, we have

$$L_\varsigma := \frac{1}{k_r d \cos\phi_\varsigma} \tag{51}$$

and $\boldsymbol{S}$ is the *"mode operator"*, which is defined in [6, 20] as

$$\boldsymbol{S} := \lim_{\varepsilon\to 0}\left( \sum_{\varsigma=-\infty}^{\infty} - \int_{-\infty}^{\infty} d\varsigma \right) e^{i\varepsilon k_r \cos\phi_\varsigma} \tag{51}$$

**Theorem 2:** *The relationship between the transverse-electric multiple scattering amplitudes of an infinite grating of dielectric circular cylinders (associated with its scattered electric and magnetic fields) for obliquely incident and horizontally polarized plane waves satisfies the following vector-matrix functional equation in which the multiple scattering amplitudes of the infinite grating* *associated with the scattered electric and magnetic fields appear in coupled form. Moreover, the single-scattering amplitudes associated with the scattered electric and magnetic fields of an isolated infinitely long cylinder corresponding to both "vertically and horizontally polarized obliquely incident waves" arise together in coupled form in this functional vector equation.*

$$\widetilde{\mathbf{G}}_z^{(TE)}(\phi,\psi_i) = \widetilde{\mathbf{g}}_\mathbf{z}(\varphi,\psi_\mathbf{i})\mathbf{1}_\mathbf{g}^{(\downarrow)} + \mathcal{S}L_\varsigma\left[\widetilde{\mathbf{g}}_\mathbf{z}(\varphi,\varphi_\varsigma)\widetilde{\mathbf{G}}_z^{(TE)}(\phi_\varsigma,\psi_i) + \widetilde{\mathbf{g}}_\mathbf{z}(\varphi,\pi-\varphi_\varsigma)\widetilde{\mathbf{G}}_z^{(TE)}(\pi-\phi_\varsigma,\psi_i)\right] \quad (53)$$

This the generalization of the Twersky's functional equation [4] for the z-components of the multiple scattering amplitudes of the electric and magnetic fields of the infinite grating of dielectric circular cylinders associated with *"horizontally polarized"* obliquely incident waves, which can be written explicitly as

$$\begin{pmatrix}\widetilde{G}_z^{(E,TE)}(\phi,\psi_i)\\ \widetilde{G}_z^{(H,TE)}(\phi,\psi_i)\end{pmatrix} = \begin{pmatrix}-g_\gamma^{\mu\varepsilon}(\phi,\psi_i)\\ g_\Gamma^{\mu\varepsilon}(\phi,\psi_i)\end{pmatrix} + \mathcal{S}L_\varsigma\begin{pmatrix}g_\Gamma^{\varepsilon\mu}(\phi,\phi_\varsigma) & -g_\gamma^{\mu\varepsilon}(\phi,\phi_\varsigma)\\ +g_\gamma^{\varepsilon\mu}(\phi,\phi_\varsigma) & g_\Gamma^{\mu\varepsilon}(\phi,\phi_\varsigma)\end{pmatrix}\begin{pmatrix}\widetilde{G}_z^{(E,TE)}(\phi_\varsigma,\psi_i)\\ \widetilde{G}_z^{(H,TE)}(\phi_\varsigma,\psi_i)\end{pmatrix}$$

$$+\mathcal{S}L_\varsigma\begin{pmatrix}g_\Gamma^{\varepsilon\mu}(\phi,\pi-\phi_\varsigma) & -g_\gamma^{\mu\varepsilon}(\phi,\pi-\phi_\varsigma)\\ +g_\gamma^{\varepsilon\mu}(\phi,\pi-\phi_\varsigma) & g_\Gamma^{\mu\varepsilon}(\phi,\pi-\phi_\varsigma)\end{pmatrix}\begin{pmatrix}\widetilde{G}_z^{(E,TE)}(\pi-\phi_\varsigma,\psi_i)\\ \widetilde{G}_z^{(H,TE)}(\pi-\phi_\varsigma,\psi_i)\end{pmatrix} \quad (54)$$

*4.3. The generalized functional matrix equation for the multiple scattering amplitudes of the infinite grating at oblique incidence*

In order to be able to express a unique functional equation describing the relationship among the multiple scattering amplitudes associated with the scattered electric and magnetic fields of the infinite grating corresponding to both vertically and horizontally polarized obliquely incident waves, we have introduced the *"matrix of multiple scattering amplitudes for the infinite grating at oblique incidence"*, namely $\widetilde{\mathbf{G}}_\mathbf{z}(\varphi,\psi_\mathbf{i})$, as

$$\left(\widetilde{\mathbf{G}}_z^{(TM)}(\phi,\psi_i) \quad \widetilde{\mathbf{G}}_z^{(TE)}(\phi,\psi_i)\right) := \widetilde{\mathbf{G}}_\mathbf{z}(\varphi,\psi_\mathbf{i}) \equiv \sum_{n=-\infty}^{\infty}\widetilde{\mathbf{A}}_\mathbf{n}(\psi_\mathbf{i})e^{in\phi} \quad (55)$$

Employing the definition above, the statements of *theorem 1* for the *"vertically polarized waves"* and *theorem 2* for the *"horizontally polarized waves"* can be combined into a unique one to implement the generalized functional matrix equation for the multiple scattering amplitudes of the infinite grating for both vertically and horizontally polarized obliquely incident waves associated with the scattered electric and magnetic fields of the infinite grating in matrix form as

$$\left(\widetilde{\mathbf{G}}_z^{(TM)}(\phi,\psi_i) \quad \widetilde{\mathbf{G}}_z^{(TE)}(\phi,\psi_i)\right) = \widetilde{\mathbf{g}}_\mathbf{z}(\varphi,\psi_\mathbf{i})\left(\mathbf{1}_\mathbf{g}^{(\uparrow)} \quad \mathbf{1}_\mathbf{g}^{(\downarrow)}\right)$$

$$+\mathbf{S}\, L_{\varsigma}\left[\tilde{\mathbf{g}}_{\mathbf{z}}(\boldsymbol{\varphi},\boldsymbol{\varphi}_{\varsigma})\left(\tilde{\mathbf{G}}_{z}^{(TM)}(\phi_{\varsigma},\psi_{i}) \quad \tilde{\mathbf{G}}_{z}^{(TE)}(\phi_{\varsigma},\psi_{i})\right)\right] \tag{56}$$

$$+\mathbf{S}\, L_{\varsigma}\left[\tilde{\mathbf{g}}_{\mathbf{z}}(\boldsymbol{\varphi},\boldsymbol{\pi}-\boldsymbol{\varphi}_{\varsigma})\left(\tilde{\mathbf{G}}_{z}^{(TM)}(\pi-\phi_{\varsigma},\psi_{i}) \quad \tilde{\mathbf{G}}_{z}^{(TE)}(\pi-\phi_{\varsigma},\psi_{i})\right)\right]$$

Exploiting (55) in the above, we have equivalently acquired the functional matrix equation for the multiple scattering amplitude of the infinite grating at oblique incidence as

$$\tilde{\mathbf{G}}_{\mathbf{z}}(\boldsymbol{\varphi},\boldsymbol{\psi}_{\mathbf{i}})=\tilde{\mathbf{g}}_{\mathbf{z}}(\boldsymbol{\varphi},\boldsymbol{\psi}_{\mathbf{i}}) \tag{57}$$

$$+\mathbf{S}\, L_{\varsigma}\left[\tilde{\mathbf{g}}_{\mathbf{z}}(\boldsymbol{\varphi},\boldsymbol{\varphi}_{\varsigma})\,\tilde{\mathbf{G}}_{\mathbf{z}}(\boldsymbol{\varphi}_{\varsigma},\boldsymbol{\psi}_{\mathbf{i}})+\tilde{\mathbf{g}}_{\mathbf{z}}(\boldsymbol{\varphi},\boldsymbol{\pi}-\boldsymbol{\varphi}_{\varsigma})\,\tilde{\mathbf{G}}_{\mathbf{z}}(\boldsymbol{\pi}-\boldsymbol{\varphi}_{\varsigma},\boldsymbol{\psi}_{\mathbf{i}})\right]$$

This can explicitly be written as

$$\begin{pmatrix}\tilde{G}_{z}^{(E,TM)}(\phi,\psi_{i}) & \tilde{G}_{z}^{(E,TE)}(\phi,\psi_{i})\\ \tilde{G}_{z}^{(H,TM)}(\phi,\psi_{i}) & \tilde{G}_{z}^{(H,TE)}(\phi,\psi_{i})\end{pmatrix}=\begin{pmatrix}g_{\Gamma}^{\varepsilon\mu}(\phi,\psi_{i}) & -g_{\gamma}^{\mu\varepsilon}(\phi,\psi_{i})\\ +g_{\gamma}^{\varepsilon\mu}(\phi,\psi_{i}) & g_{\Gamma}^{\mu\varepsilon}(\phi,\psi_{i})\end{pmatrix}$$

$$+\mathbf{S}\left\{L_{\varsigma}\left[\begin{pmatrix}g_{\Gamma}^{\varepsilon\mu}(\phi,\phi_{\varsigma}) & -g_{\gamma}^{\mu\varepsilon}(\phi,\phi_{\varsigma})\\ +g_{\gamma}^{\varepsilon\mu}(\phi,\phi_{\varsigma}) & g_{\Gamma}^{\mu\varepsilon}(\phi,\phi_{\varsigma})\end{pmatrix}\begin{pmatrix}\tilde{G}_{z}^{(E,TM)}(\phi_{\varsigma},\psi_{i}) & \tilde{G}_{z}^{(E,TE)}(\phi_{\varsigma},\psi_{i})\\ \tilde{G}_{z}^{(H,TM)}(\phi_{\varsigma},\psi_{i}) & \tilde{G}_{z}^{(H,TE)}(\phi_{\varsigma},\psi_{i})\end{pmatrix}\right]\right\} \tag{58}$$

$$+\mathbf{S}\left\{L_{\varsigma}\left[\begin{pmatrix}g_{\Gamma}^{\varepsilon\mu}(\phi,\pi-\phi_{\varsigma}) & -g_{\gamma}^{\mu\varepsilon}(\phi,\pi-\phi_{\varsigma})\\ +g_{\gamma}^{\varepsilon\mu}(\phi,\pi-\phi_{\varsigma}) & g_{\Gamma}^{\mu\varepsilon}(\phi,\pi-\phi_{\varsigma})\end{pmatrix}\begin{pmatrix}\tilde{G}_{z}^{(E,TM)}(\pi-\phi_{\varsigma},\psi_{i}) & \tilde{G}_{z}^{(E,TE)}(\pi-\phi_{\varsigma},\psi_{i})\\ \tilde{G}_{z}^{(H,TM)}(\pi-\phi_{\varsigma},\psi_{i}) & \tilde{G}_{z}^{(H,TE)}(\pi-\phi_{\varsigma},\psi_{i})\end{pmatrix}\right]\right\}$$

*4.4. Proof of the theorems 1 and 2 of the infinite grating at oblique incidence*

In order to prove the theorems 1 and 2, we will show that the generalized functional matrix equation in (57) is valid. For this purpose, we introduce the alternate representation of the *"Schlömilch series,* $H_{n-m}(k_r d)$ *"* for obliquely incident plane waves [20] as

$$H_{n-m}(k_r d)=\mathbf{S}\, L_{\varsigma}\left[e^{-i(n-m)\phi_{\varsigma}}+e^{-i(n-m)(\pi-\phi_{\varsigma})}\right] \tag{59}$$

which is expressed by Twersky [5] for normal incidence. In terms of $H_{n-m}$, the matrix equation (45) for the multiple scattering coefficients of the infinite grating for obliquely incident waves can be rewritten as

$$\tilde{\mathbf{A}}_{\mathbf{n}}(\boldsymbol{\psi}_{\mathbf{i}})=\tilde{\mathbf{a}}_{\mathbf{n}}\left[e^{-in\psi_{i}}+\sum_{m=-\infty}^{\infty}\tilde{\mathbf{A}}_{\mathbf{m}}(\boldsymbol{\psi}_{\mathbf{i}})H_{n-m}(k_r d)\right] \tag{60}$$

Inserting the matrix $\tilde{\mathbf{A}}_{\mathbf{n}}(\boldsymbol{\psi}_{\mathbf{i}})$ in (60) into $\tilde{\mathbf{G}}_{\mathbf{z}}(\boldsymbol{\varphi},\boldsymbol{\psi}_{\mathbf{i}})$, the *"matrix of multiple scattering amplitudes for the infinite grating at oblique incidence"*, given in expression (55), we have obtained

$$\tilde{\mathbf{G}}_{\mathbf{z}}(\boldsymbol{\varphi},\boldsymbol{\psi}_{\mathbf{i}})=\sum_{n=-\infty}^{\infty}\tilde{\mathbf{A}}_{\mathbf{n}}(\boldsymbol{\psi}_{\mathbf{i}})e^{in\phi}=\sum_{n=-\infty}^{\infty}\tilde{\mathbf{a}}_{\mathbf{n}}\left[e^{-in\psi_{i}}+\sum_{m=-\infty}^{\infty}\tilde{\mathbf{A}}_{\mathbf{m}}(\boldsymbol{\psi}_{\mathbf{i}})H_{n-m}(k_r d)\right]e^{in\phi} \tag{61}$$

or, equivalently

$$\widetilde{\mathbf{G}}_{\mathbf{z}}(\varphi,\psi_{\mathbf{i}}) = \sum_{n=-\infty}^{\infty} \widetilde{\mathbf{a}}_{\mathbf{n}} e^{in(\phi-\psi_i)} + \sum_{n=-\infty}^{\infty} \widetilde{\mathbf{a}}_{\mathbf{n}} \left[ \sum_{m=-\infty}^{\infty} \widetilde{\mathbf{A}}_{\mathbf{m}}(\psi_{\mathbf{i}}) H_{n-m}(k_r d) \right] e^{in\phi} \tag{62}$$

Recognizing the first term in the equation above as the *"normalized single scattering amplitude matrix at oblique incidence"*, viz., $\widetilde{\mathbf{g}}_{\mathbf{z}}(\varphi,\psi_{\mathbf{i}})$ we have

$$\widetilde{\mathbf{G}}_{\mathbf{z}}(\varphi,\psi_{\mathbf{i}}) = \widetilde{\mathbf{g}}_{\mathbf{z}}(\varphi,\psi_{\mathbf{i}}) + \sum_{n=-\infty}^{\infty} \widetilde{\mathbf{a}}_{\mathbf{n}}\, e^{in\phi} \left[ \sum_{m=-\infty}^{\infty} \widetilde{\mathbf{A}}_{\mathbf{m}}(\psi_{\mathbf{i}}) H_{n-m}(k_r d) \right] \tag{63}$$

Introducing the Schlömilch series $H_{n-m}(k_r d)$ in (59) for obliquely incident plane waves into the expression (63), we have

$$\widetilde{\mathbf{G}}_{\mathbf{z}}(\varphi,\psi_{\mathbf{i}}) = \widetilde{\mathbf{g}}_{\mathbf{z}}(\varphi,\psi_{\mathbf{i}})$$

$$+ \sum_{n=-\infty}^{\infty} \widetilde{\mathbf{a}}_{\mathbf{n}}\, e^{in\phi} \left\{ \sum_{m=-\infty}^{\infty} \widetilde{\mathbf{A}}_{\mathbf{m}}(\psi_{\mathbf{i}}) \mathbf{S}\, L_{\varsigma} \left[ e^{-i(n-m)\phi_{\varsigma}} + e^{-i(n-m)(\pi-\phi_{\varsigma})} \right] \right\} \tag{64a}$$

$$\widetilde{\mathbf{G}}_{\mathbf{z}}(\varphi,\psi_{\mathbf{i}}) = \widetilde{\mathbf{g}}_{\mathbf{z}}(\varphi,\psi_{\mathbf{i}}) + \mathbf{S}\, L_{\varsigma} \left\{ \sum_{n=-\infty}^{\infty} \widetilde{\mathbf{a}}_{\mathbf{n}}\, e^{in\phi} \left( \sum_{m=-\infty}^{\infty} \widetilde{\mathbf{A}}_{\mathbf{m}}(\psi_{\mathbf{i}})\, e^{i(m-n)\phi_{\varsigma}} \right) \right\}$$

$$+ \mathbf{S}\, L_{\varsigma} \left\{ \sum_{n=-\infty}^{\infty} \widetilde{\mathbf{a}}_{\mathbf{n}}\, e^{in\phi} \left( \sum_{m=-\infty}^{\infty} \widetilde{\mathbf{A}}_{\mathbf{m}}(\psi_{\mathbf{i}})\, e^{i(m-n)(\pi-\phi_{\varsigma})} \right) \right\} \tag{64b}$$

Interchanging the order of summation, we have acquired the matrix of the multiple scattered amplitudes of the infinite grating for obliquely incident waves associated with both polarizations as

$$\widetilde{\mathbf{G}}_{\mathbf{z}}(\varphi,\psi_{\mathbf{i}}) = \widetilde{\mathbf{g}}_{\mathbf{z}}(\varphi,\psi_{\mathbf{i}}) + \mathbf{S}\, L_{\varsigma} \left\{ \sum_{n=-\infty}^{\infty} \widetilde{\mathbf{a}}_{\mathbf{n}}\, e^{in(\phi-\phi_{\varsigma})} \left( \sum_{m=-\infty}^{\infty} \widetilde{\mathbf{A}}_{\mathbf{m}}(\psi_{\mathbf{i}})\, e^{im\phi_{\varsigma}} \right) \right\}$$

$$+ \mathbf{S}\, L_{\varsigma} \left\{ \sum_{n=-\infty}^{\infty} \widetilde{\mathbf{a}}_{\mathbf{n}}\, e^{in[\phi-(\pi-\phi_{\varsigma})]} \left( \sum_{m=-\infty}^{\infty} \widetilde{\mathbf{A}}_{\mathbf{m}}(\psi_{\mathbf{i}})\, e^{im(\pi-\phi_{\varsigma})} \right) \right\} \tag{65}$$

Employing the definitions of the single and multiple scattered amplitudes in the expression (47) and (55) for the following terms

$$\widetilde{\mathbf{g}}_{\mathbf{z}}(\varphi,\varphi_{\varsigma}) = \sum_{n=-\infty}^{\infty} \widetilde{\mathbf{a}}_{\mathbf{n}}\, e^{in(\phi-\phi_{\varsigma})} \tag{66a}$$

$$\widetilde{\mathbf{G}}_{\mathbf{z}}(\varphi_{\varsigma},\psi_{\mathbf{i}}) = \sum_{m=-\infty}^{\infty} \widetilde{\mathbf{A}}_{\mathbf{m}}(\psi_{\mathbf{i}})\, e^{im\phi_{\varsigma}} \tag{66b}$$

$$\widetilde{\mathbf{g}}_{\mathbf{z}}(\varphi,\pi-\varphi_{\varsigma}) = \sum_{n=-\infty}^{\infty} \widetilde{\mathbf{a}}_{\mathbf{n}}\, e^{in\left[\phi-(\pi-\phi_{\varsigma})\right]} \tag{66c}$$

$$\widetilde{\mathbf{G}}_{\mathbf{z}}(\pi-\varphi_{\varsigma},\psi_{\mathbf{i}}) = \sum_{m=-\infty}^{\infty} \widetilde{\mathbf{A}}_{\mathbf{m}}(\psi_{\mathbf{i}})\, e^{im(\pi-\phi_{\varsigma})} \tag{66d}$$

Inserting these definitions to the equation (65), we have finally acquired the *"functional matrix equation for the multiple scattered amplitudes of the infinite grating at oblique incidence"* as it was anticipated by the statements of theorems 1 and 2 in the expression (57). The scalar version of this grating equation was originally derived by Twersky [4] for *"normally incident waves"*. Thus, we have proved that the functional relationship among the multiple scattered amplitudes of an infinite grating for obliquely incident waves obey to the matrix equation given in (57). As a consequence of this, the functional matrix equation in (57) comprise the vector functional equation of theorems 1 and 2, and the statements of those theorems are valid.

## 5. Conclusion

The rigorous analytical representation for both transverse magnetic (TM) and transverse electric (TE) multiple scattering coefficients of the fields radiated by an infinite grating of cylinders excited by an obliquely incident plane wave is generated by the application of *"the direct Neumann iteration procedure"* to two infinite sets of equations describing the mutual coupling relationship among the multiple scattering coefficients of the infinite grating. The exact solution is expressed in terms of the well-known *"single scattering coefficients of an isolated dielectric circular cylinder at oblique incidence",* which was originally derived by Wait [29], and *"Schlömilch series"* [6]. In this study, we have acquired the analytic expressions up to the fourth order of scattering by iteration for the multiple scattering coefficients associated with both the transverse magnetic (TM) and transverse electric (TE) modes of the infinite grating for obliquely incident waves. It has been proved that the exact representations for both TM and TE multiple scattering coefficients of the infinite grating at oblique incidence comprise *"TM and TE scattering coefficients of an isolated cylinder for obliquely incident waves together in coupled form"* as well as *"Schlömilch series",* which ensure the coupling between TM and TE modes. The exact solution for the multiple scattering coefficients of the infinite grating acquired in this investigation takes into account all possible contributions to the excitation of a particular cylinder by the radiation scattered by the remaining cylinders at oblique incidence. In addition, it reduces to the multiple scattering coefficients of the infinite grating at non-oblique incidence obtained by Twersky [5] as the oblique angle of arrival *"* $\theta_i$ *"* made with $z$-axis approaches to *"* $\pi/2$ *(normal incidence)"*, while it represents the scattering coefficients of an isolated cylinder at oblique incidence derived originally by Wait [17] as the distance *"d"* between the individual cylinders of the grating approaches to infinity. Furthermore, the generalization of the sum-integral grating equation of Twersky [4] is obtained for the *"matrix of normalized multiple scattered amplitudes* $\widetilde{\mathbf{G}}_{\mathbf{z}}(\varphi_\varsigma, \psi_i)$ *of a cylinder at oblique incidence in the grating"* for the direction of incidence angle $\psi_i$ (made with $x$-axis) and observation angle $\varphi_\varsigma$ (made with $x$-axis), in terms of the mode operator $\boldsymbol{S} := \lim_{\varepsilon\to 0}\left(\sum_{\varsigma=-\infty}^{+\infty} - \int_{-\infty}^{+\infty} d\varsigma\right) e^{i\varepsilon k_r \cos\phi_\varsigma}$ and the *"matrix of normalized scattering amplitudes* $\widetilde{\mathbf{g}}_{\mathbf{z}}(\varphi, \psi_i)$ *of an isolated cylinder at oblique incidence by an angle of obliquity* $\theta_i$ (made with z-axis),", as $\widetilde{\mathbf{G}}_{\mathbf{z}}(\varphi, \psi_i) = \widetilde{\mathbf{g}}_{\mathbf{z}}(\varphi, \psi_i) + \boldsymbol{S}\, L_\varsigma \left[\widetilde{\mathbf{g}}_{\mathbf{z}}(\varphi, \varphi_\varsigma)\, \widetilde{\mathbf{G}}_{\mathbf{z}}(\varphi_\varsigma, \psi_i) + \widetilde{\mathbf{g}}_{\mathbf{z}}(\varphi, \pi - \varphi_\varsigma)\, \widetilde{\mathbf{G}}_{\mathbf{z}}(\pi - \varphi_\varsigma, \psi_i)\right]$ where $L_\varsigma$ s are some known constants.

(ÖK) DIVISION OF ELECTROPHYSICS RESEARCH, FACULTY OF COMPUTER SCIENCES, IZMIR UNIVERSITY OF ECONOMICS, IZMIR 35330 TURKEY.
E-mail address: omer.kavaklioglu@ieu.edu.tr; omer_kavaklioglu@yahoo.com

(BS) DEPARTMENT OF MATHEMATICS, FACULTY OF SCIENCES AND LITERATURE, IZMIR UNIVERSITY OF ECONOMICS, IZMIR 35330 TURKEY.
E-mail address: baruch.schneider@ieu.edu.tr